\documentclass[preprint]{aastex}

\usepackage{graphicx}
\usepackage{epsf}

\def\ni{\noindent}

\def\beq{\begin{equation}}
\def\ee{\end{equation}}
\def\lsim{\mathrel{\rlap{\lower4pt\hbox{\hskip1pt$\sim$}}
    \raise1pt\hbox{$<$}}}
\def\gsim{\mathrel{\rlap{\lower4pt\hbox{\hskip1pt$\sim$}}
    \raise1pt\hbox{$>$}}}

\def\lb{\langle}
\def\rb{\rangle}

\begin{document}

\title{
Incorporating Human Body Mass in Standards of Helmet Impact Protection
against Traumatic Brain Injury}

\author{Eric G. Blackman$^{1,2}$}
 
\affil{1. Dept. of Physics and Astronomy, Univ. of Rochester, Rochester, NY 14627, USA}

\setcounter{equation}{0}
\begin{abstract}

Impact induced traumatic brain injury (ITBI) describes  brain injury  from head impact  not necessarily accompanied by skull fracture.  For sufficiently abrupt head impact decelerations, ITBI results from brain tissue stress incurred as the brain crashes into the inside of the skull wall, displacing the surrounding  cerebral spinal fluid (CSF).  Proper helmet cushioning can damp the impact  force and reduce ITBI.  But force is mass times acceleration and  current helmet blunt impact standards are based only on acceleration thresholds. Here I show how this implies that present  standards grossly overestimate the minimum  acceleration onset for ITBI by implicitly assuming that the brain is mechanically decoupled from the body.  I quantify how an arbitrary   orientation of the body with respect to  impact direction increases the effective mass that should be used in calculating the required damping force and injury threshold accelerations. I suggest a practical method to incorporate the body mass and impact angle into ITBI helmet standards and point out directions for further work.

\end{abstract}

\section{Introduction}

Traumatic Brain Injury (TBI) refers to physical injury to the brain  not necessarily  accompanied by visible external head injury (e.g. Bandak et al. 1996).  Impact induced TBI (ITBI) arises from
rapid acceleration or deceleration of the head, as can occur in sports,
motor vehicle accidents, or in military combat. 
The brain is composed of soft tissue  and is surrounded by a layer of cerebral spinal 
fluid  (CSF) inside the head. 
During  normal head motions,  the force on the head is small enough that the 
CSF prevents  the head from impacting the skull wall.
In contrast,  the magnitude of deceleration upon impact 
is so large that the CSF  cannot adequately protect the brain. As the skull comes to a stop, the brain  pummels the the inner skull pushing the fluid away.  The brain deformation can occur fast enough to leave a small cavity between skull and brain at the antipode.  
As the brain rebounds it slaps into the CSF such that countre-coup injury can occur.   
The brain deformation can also induce more diffuse brain tissue injury via shear stress.

The medical consequences of such ITBI  range from minor concussions to 
complete cognitive impairment.  
 Almost 2 million civilian cases of TBI (virtually all are ITBI)
 have been diagnosed each year since the early 1990s, leading to $ ~200$ 
hospitalizations per 100,000 people each year and 56,000 deaths per year
(McArthur et al. 2004).
About 50\% of TBI cases come from automobile accidents
and 20\% from sports related injuries (Bohnen et al. 1992). 
The majority of TBI are classified as mild traumatic
brain injury (MTBI) or concussion.  
TBI is also common among  military combat personnel (Okie 2005). 
Recent estimates suggest as high as $\sim 20$\% of US soliders returning from Iraq and Afghanistan have incurred TBI
(Terrio et al. 2009). In this context, 
 TBI is likely a combination of ITBI and blast induced TBI (BTBI).
the latter referring to the direct effect of 
blast overpressure (Cernak  2005; Taber et al. 2006;  Moss et al. 2009)
 which adds to any additional ITBI.

Protection against ITBI requires helmets with proper cushioning and a proper blunt impact measure to determine the effectiveness of such helmets. 
Current helmet blunt impact standards are derived from empirically determined injury measures
of acceleration vs. duration based on cadaver and scaled monkey data (Ono et al. 1980).  
Drop tests of helmeted head forms fitted with accelerometers for chosen drop heights
then empirically test whether a given helmet falls within the acceptable 
acceleration range upon  impact (e.g. McEntire et al. 2005).  But force equals mass times acceleration,
so the use of acceleration thresholds without incorporating head and body mass  is flawed. 
Using only the head form + helmet mass may be appropriate for computing the impact force on the head for a body oriented perpendicular to the direction of impact, but the effective mass increases for impacts with the body increasingly aligned with the direction of impact
 because some fraction of the force incurred by the body is transmitted through the skull to he brain.   For exact alignment, the force would depend on the entire body mass.


In section 2, I give a simple derivation of the physics principles behind 
helmet protection to blunt impact and ITBI. In section 3, I discuss
the quantitative flaws of current blunt impact/TBI standards.
In section 4, I derive corrections to standard threshold TBI measures
that incorporates the body impact angle and thus the effective mass of 
impact.  In section 5 I describe how these corrections can be implemented 
in practice in future work and conclude in Sec 6.

\section {Why Cushioning Reduces Impact Force}

Newton's equation of motion for an  object of mass $m$ 
subject to  a force is 
\beq
m{d{\bf v}\over dt}={\bf F}= m{\bf a},
\label{1}
\ee
where ${\bf F}$ is the force, and ${\bf v}$ is the speed.
and ${\bf a}$ is the acceleration.
 For an object incurring  a drop and impact,   Eq. (\ref{1}) is used to compute the motion
during  free fall, 
 and the deceleration
 upon impact determined by the helmet properties.
 Using  ${\bf a}=(0,0,a)$ and ${\bf v}=(0,0,v)$ 
(i.e. both with only  $z$ components)
and assuming that $|{v \over dv/dt}| << |{a\over da/dt}|$
we can integrate (\ref{1}) equation to obtain 
\beq
v(t)= v_0+ a t,
\label{2}
\ee
where $v_0$ is the initial speed at initial height $z_0$.
Integrating (\ref{2}) gives 
\beq
z(t)= z_0 + v_0t + {1\over 2} a t^2.
\label{3}
\ee
Eliminating $t$ from (\ref{2}) and (\ref{3})
gives 
\beq
v^2(t)- v_0^2 = 2a d,  
\label{4}
\ee
where $d= z(t)-z_0$.

We can use (\ref{4}) to compute the maximum free fall speed
reached just before impact for an object dropped from rest at height $h$
when $a$ is given by Earth's gravitational acceleration $g
=-10 {\rm m/s^2}=-32{\rm ft/s^2}$.
For $v_0=0$, $z_0=h$, and $z(t_I)=0$,   
Eq. (\ref{4}) implies the speed toward the ground at the time of  impact $t_I$ is
\beq
v_I=v(t_I)=(2gh)^{1/2}.
\label{4a}
\ee 
This gain in speed 
corresponds to gain kinetic energy at the expense of 
 gravitational potential energy.  
Upon impact, most of  
this kinetic energy is converted into work done
in deforming and stopping the object. This work 
can be expressed as the force incurred times the stopping distance,
and equals the kinetic energy just before impact. That is, 
\beq
F_s s={1\over 2}m v_I^2,
\label{5a}
\ee
where $F_s=ma_s$ is the force exerted on the object by the stopping 
acceleration $a_s$ upon impact over the stopping distance $s$.
Combining  Eq. (\ref{4a})  and Eq. (\ref{5a}) 
gives
\beq
a_s= {{v_I}^{2}\over 2s}={g h\over s},
\label{6}
\ee
showing that increasing $s$  reduces the magnitude of acceleration and thus force of impact
(see also Cory et al. 2002).

A larger  stopping distance $s$, also implies a longer
stopping time: By applying equation (\ref{4}) to 
the case in which the object's initial speed corresponds to the 
speed of impact $v_I$ from (\ref{4a}) and  taking 
the final speed $v(t> t_I)=0$  as the object comes to rest, we obtain
$v_0=(2gh)^{1/2}$. Plugging this into (\ref{3}), setting $|z(t)-z_0 |= s$, and using (\ref{6})  
we have 
\beq
s=(2gh)^{1/2}t+  {gh\over2 s}t^2.
\label{6a}
\ee
Solving (\ref{6a}) for $t>0$ gives 
\beq
t=(2-{\sqrt 2}){s\over (gh)^{1/2}}
\label{6a}
\ee
which highlights that the longer the stopping distance, $s$, 
the longer the deceleration time $t$  for an object
that acquired its impact speed by falling from height $h$.

Eqs. 
(\ref{6}) and (\ref{6a})
show that  if cushioning can  increasing the distance or time 
over which a headform decelerates from its maximum speed to zero, the magnitude of
acceleration  of the impacting object  is reduced, and thus so is the force of impact.

 The amount of tissue damage and TBI
 depends on a  combination of the external 
force and the time scale over which the force
acts. Below some minimum threshold force, determined by the 
biological tissue properties, no damage will occur no matter
how long the force is applied. However, 
a small force above this threshold  acting 
over a long time could do more damage that a much larger
force over a short time.  For a given mass of impactor, 
 empirically determined  damage curves, in principle, 
provide a practical method for 
identifying an injury threshold curve in the force vs. duration plane.
As I describe in the next section,  present curves 
are construted in the acceleration vs. duration plane, 
and practical  application of these curves has fundamental shortcomings.

\section{Shortcomings of Current Head Impact TBI Protection Indices}

The peak acceleration incurred for a fixed mass impactor 
indicates the peak force providing one measure of potential injury.
However, the  need to incorporate 
a combination of acceleration and duration into an  
injury measure (see Hayes et al. 2007 for  review)
 was evident from 
the Wayne State Tolerance Curve (WTSC) (Pattrick et al. 1963, Snyder 1970) supposedly be tolerated without severe head injury (considered to be skull fracture).  The  original data  came from  (1) drop tests of 4 embalmed cadaver heads on plates, with measurements of linear acceleration, intracranial pressure and skull damage  (2) air blasts to exposed cadaver brains and (3) hammer blows to animals. The data showed that small
accelerations can be tolerated for longer durations than large accelerations.
 The
severity index (SI)  (Gadd 1966)
quantifies the WSTC into a (unfortunately dimensional) quantity given by 
\beq
SI \equiv \int_{t_1}^{t_2} a_g(t)^{5/2} dt 
\ee
where $a_g$ is the dimensionless 
acceleration in units of gravity
$g$ and $t$ is measured in seconds. 

The SI incorrectly implies that  impacts of extremely slow deceleration extended over a very long period give the same injury threshold as  high deceleration impacts of very short duration, whereas there is  no injury at very
low accelerations. 
This is partly corrected by the Head Injury Criterion (HIC) (Versace 1971)
\beq
HIC=\left[(t_2-t_1)\left\{ {1\over t_2-t_1} \int^{t_2}_{t_1} a_g dt \right\}^{5/2}
\right]_{max},
\ee
which restricts the  SI to an  integral near an empirically 
chosen time interval (measured in seconds)
   $t_2-t_1$ near the peak acceleration.
 When the acceleration magnitude is nearly constant over the chosen time interval, $HIC \simeq SI \propto a_g^{2.5}$.
Typically an HIC between 500 and 2500 
is converted into a probability for fatality or concussion.  
As applied to non-fatal TBI, 
Ono et al. (1980) performed experiments both with  human cadavers and live monkeys and determined distinct human thresholds of skull fracture vs. concussion, a form of TBI.  The latter TBI threshold curve has been called the Japanese Head Tolerance Curve (JHTC).

But the SI and HIC  standards  are flawed.
Note for example, that  concussions in the NFL
are occuring with  higher probablity that the JHTC curve would predict
at measured values of the head acceleration (Viano et al. 2006) .
In addition,  King et al. (2000, 2003) and Zhang et al. (2004) used video footage of helmet-helmet collisions in games from the National Football League (NFL) in which known concussions occurred. The motions producing the concussions were reproduced in the laboratory using helmeted dummies, with linear and rotational accelerometers. The data measured were then fed as initial conditions into head impact computer simulations that include a comprehensive computer model of the human head and brain (Wayne Stead Head Model).   By analyzing the stresses on the simulated brain tissue, 
the HIC proved to be no better than the peak linear acceleration, or the head impact power (HIP, Newman et al. 2000), an uncommonly used 
measure of the total kinetic energy per unit time.
The HIP was marginally the best correlator, followed by the peak acceleration and then the HIC, and rotational acceleration.  

In principle, the conceptual advantage of the 
HIP would be that it includes mass whereas the SI, HIC and peak accelerations do not include the mass. However, typically the mass is used is that for the head itself not adjusted for impact angle and body mass.  Also impacts analyzed from drop tests in the laboratory 
use approximately the same mass of head forms and helmets, so the relative change in effective mass as a function of body impact angle is not present. 
The non-inclusion of the mass is a  conceptual shortcoming that  I quantify in the next section.

\section{Incoporating Impact  Area and Body Mass into TBI Protection Standards}


Mechanical stress on brain tissue causes  TBI and how external impact forces
produce specific clinical manifestations of ITBI comprises a
 complex set of questions.   But the role of helmet 
protection  is largely independent of the specific TBI manifestation:  
a helmet accomplishes much by simply reducing the overall stress on brain tissue. 
For a given acceleration  and fixed mass, 
the local  stress  is reduced for a larger brain surface area of impact.  
For a given surface area and a given acceleration,
an increased mass will produce more force per unit area and thus more stress. 

Using the reasonable assumption that that material threshold for brain tissue damage is the same in woodpeckers and humans,
Gibson (2006) showed  why woodpeckers would not be expected to get concussions even though they incur high enough accelerations over long enough durations to
exceed the TBI threshold of the JHTC curve.  
The material stress associated with the very
high HIC value is still below the  tissue damage threshold when applied to a 
woodpecker head. The same HIC value corresponds to  a much higher force per unit area when applied to the human head. 

Complementarily, when comparing  impacts of the same 
surface area but different effective masses, a single HIC standard is also
inadequate because force per unit area depends on mass.  
A person oriented  vertically during a fall on their head will incur more head force  compared to a person oriented horizonally during the fall.
The stress incurred by the brain as it contacts the inner skull during the head impact is a combination of (1) the force need to stop the brain as if it were isolated in free fall, plus (2)  a contribution that comes from stress waves propagating into the brain from the skull 
 which are  sourced by the weight of the entire body which is coupled to the brain via the CSF.
The force on the skull depends on the  mass aligned along the direction of impact, but because the coupling between the skull and brain is likely less than  100\% efficient, 
there is some efficiency coefficient that scales the force on the skull to that on the brain for a fixed effective mass along the direction of impact. 
In a more detailed study, it may turn out that this coefficient depends on mass but here  I consider the simple case in which it does not, and then take ratios of quantities in which the
coefficient  cancles out.


 
To see the role of the body mass quantitatively, 
let  $0\le\theta\le \pi/2$ be the  impact angle between the 
line passing through the body center of mass and impact point,
and the line  through the impact point in the direction of center of mass momentum before impact.
The case  $\theta=\pi/2$ correspsonds to the body oriented horizontally for a vertical 
fall impact and $\theta=0$ corresponds to the body oriented vertically for a vertical impact.
For any $0\le \theta < \pi/2$, 
the effective mass of the head will be larger than just that of the head form.

Consider two cases labeled by 1 and 2, which respectively produce 
a force per unit area of $\sigma_{1}$ and $\sigma_{2}$ on the brain.
Let 
$\sigma_c$ be a property of brain tissue indicating the threshold stress above which TBI occurs.  Taking
$\sigma_1=\sigma_c=\sigma_2$, and expressing 
  $\sigma_1/\sigma_2$, in terms of the separate properties of each
system, we have
\beq
{\sigma_{1}\over \sigma_{2}} = 1= {F_1 A_1 \over F_2 A_2}
={[(m_{b1}-m_{h1})cos\theta_1 + m_{h1}]a_1 A_1
\over [(m_{b2}-m_{h2})cos\theta_2 + m_{h2}]a_2 A_2
},
\label{gib}
\ee
where $F_1$ and $F_2$ are the forces on the respective heads during impact deceleration; $A_1$ and $A_2$ are the head contact areas; 
$m_{h1},m_{h2}$ and $m_{b1},m_{b2}$ are the respective head and total 
body  masses for the two cases, 
$\theta_1$, $\theta_2$ are the respective impact angles, and $a_1,a_2$ are
the magnitudes of the deceleration from maximum to zero upon impact.
If  the two impacting bodies are identical 
but differ in impact angles for the cases considered, we can 
set  $A_1=A_2$, $m_{h1}=m_{h2}=m_h$, 
$m_{b1}=m_{b2}=m_b$ in (\ref{gib}). After a bit of algebra, this gives 
\beq
{a_1\over a_2}=
{ (m_b/m_h-1)cos\theta_2 + 1 \over 
(m_{b}/m_h-1)cos\theta_1 + 1}.
\ee
If we take $\theta_2=\pi/2$ as a fiducial baseline case 
corresponding to the body perpendicular to the direction of impact, we obtain
\beq
{a_1\over a_2}=
{ 1 \over 
(m_{b}/m_h-1)cos\theta_1 + 1}.
\label{14}
\ee
This  formula is plotted in Fig. 1a. For 
a fixed head+helmet mass of $m_h=6.4$kg,
the three curves in the figure  correspond to  body masses of $m_b=64,\ 82,\ 100$kg respectively.
Each point on these curves corresponds
to the impact force imparting the same TBI threshold stress.
 The curves show that this stress  arises
for  a significantly lower magnitude of head aceleration
when the body angle of impact deviates from the fiducial angle of $\pi/2$.


For most arenas of injury (e.g. football, military, motor vehicle accidents)
the angle of impact will vary from incident to incident so
 practical incorporation of the effect of angle into a TBI standard 
requires either a 
a conservative standard that protects for impact angles down to a 
chosen minimum $\theta_{min}$, or  a suitable average over a range of angles.
For the latter, 
 the average of (\ref{14})  in spherical polar coordinates is
\beq
{\left\lb a_1\over a_2\right\rb}=
{\int^{\mu_{1,min'}}_{0}
{1 \over 
(m_{b}/m_h-1)\mu'_1 + 1}d \mu'_1\over 
\int^{\mu_m'}_{0}d\mu_1'}
\label{15},
\ee
where $\mu_{1,min}'=cos\theta'_{1,min}$, the cosine of the minimum impact angle
(where $cos \theta_{1}=0$ corresponds to impact direction perpendicular to body alignment and $cos \theta_{1}=1$ corresponds to body alinged parallel
to direction of impact).  
Fig. 1b  shows plot the average of 
(\ref{14}) over $\theta_{min} \le \theta \le \pi/2$ as a
function of the choice of $\theta_{min}$.

As dicsussed in Sec 2., for a nearly constant acceleration over  impact duration, the 
HIC and the SI  are proportional to $a^{2.5}$. By analogy to 
(\ref{14}) and (\ref{15}) we can then write
\beq
{ HIC_{1}\over HIC_2 }
\simeq{ SI_{1}\over SI_2 }\simeq
{a_1^{2.5}\over a_2^{2.5}}=
{1 \over 
(m_{b}/m_h-1)\mu'_1 + 1}
\label{16},
\ee
and for the average
\beq
{\left\lb HIC_{1}\over HIC_2 \right\rb}\simeq
{\left\lb a_1^{2.5}\over a_2^{2.5}\right\rb}=
{\int^{\mu_{1,min}'}_{0}
\left({1 \over 
(m_{b}/m_h-1)\mu'_1 + 1}\right)^{2.5}d \mu'_1\over 
\int^{\mu_{m}'}_{0}d\mu_1'}
\label{17}.
\ee
Eqs. (\ref{16}) and (\ref{17}) are plotted in the bottom row of  Fig.1  for
mass ratios  ${m_b\over m_h}=10, 12.81, 15.63$ 
corresponding to the bottom, middle, and top curves respectively in each panel.

\section{Prescription for Revising ITBI Helmet Standards}

The  shapes of all  curves in Fig 1.  flatten at small $\theta$ and steepen 
near $\theta =1.1$ ($\sim 63$ degrees).
Thus for either row 1 (peak acceleration) or row 
 2 (HIC), there is 
 is a dramatic drop in the critical  thresholds for injury
even for angles that deviate only $\sim 30\%$  from the fiducial 
 $\theta=\pi/2=1.57$ rad. Complementarily, the curves in the right column panels of Fig 1. 
highlight that if impact angles are quasi-random over a  range,
averaging over this range for different choices of minimum impact angle 
$\theta_{min}$ 
is relativiely insensitive to this choice for $\theta_{min} < 1.1$ rad.
Overall, the plots show that it may
not be too much more  demanding to protect against the full range of impact angles below 
$\theta < 1.1$ rad than it is to protect impact anges  $1.1<\theta \le \pi/2$ rad.

Using  the calculations and Fig. 1. a  
procedure for straightforward improvement of ITBI helmet protection standards
emerges:
(1) Identify either the peak acceleration, or the SI, or HIC index on the usual JHTC type curve corresponding to the supposed acceptable injury threshold
for the characteristic time scale characteristic of the particlar
impact (e.g. football helmet collision).
This provides the fiducial acceleration, the value corresponding to
$\theta=\pi/2$ in the calcuations above.   Assume that this threshold is correct to produce
supercritical stress on brain tissue for TBI based on acceleration of only the head form (e.g. helmet + head).
(2) Pick a standard head mass AND body mass for a standard victim
based on a practical statistical criterion of characteristic individuals involved.
(3) Choose a characteristic minumum impact angle to accomodate
a statistically significant fraction of all impacts based on  carefully assessment of the types of impacts occurred in the activity and the equivalent range of impact angles.
(4) Find the correction factor to the peak acceleration or HIC 
compared helmet drop tests in the laboratory
either from plots like the left  panels of Fig. 1 for the chosen angle, 
 or the right panels of Fig. 1. using the chosen angle as the lower bound
 for  averaging over an angular
range.

For example, consider a head impact duration to be $\sim 15$ms.  
The 30\% risk for concussions using the conventional   
JHTC corresponds to $a_1=125g$ for this duration. Let us assume that this is 
the correct threshold for TBI based on acceleration incurred
for  a helmeted head form of  $m_h\sim 6.4$kg in a drop test.
Now take $m_b=100$kg so that $m_b/m_h= 15.63$
and consider typical cofllisions to take place at an equivalent 
impact angle range between $\theta_{min}=\pi/ 4 \le \theta \le \pi/2$. 
Using the bottom curves in the plots on the right column 
panels of Fig 1., (which correspond to the chosen mass ratio),
we find a reduction  to the peak acceleration index threshold by a factor 
$1/4$, and a reduction to the HIC threshold by a factor $1/20$.

\section{Conclsuion}

Commonly used ITBI helmet protection standards are based on emprical 
injury threshold curves of acceleration vs. impact duration from motor vehicle crash studies. Presently, helmet blunt impact and ITBI protection testing typically
evaluate whether the acceleration upon impact from drop tests 
of helmet-fitted head form falls sufficiently below the injury threshold
from these curves.   I discussed that the resulting  curves from this procedure
can significantly overestimate the minimum acceleration for ITBI because
they do not take into account the body angle of impact, and thus the effective mass of impact. 
The force on the head and brain is mass times acceleration and standard measures of protection based on acceleration can at most apply to a fixed mass of impactor.
This absence of inclusion of effective mass in standard blunt impact criteria such as peak acceleration or HIC  may explain for example, why concussions are seen in NFL football
at lower peak accelerations than expected.

By incorporating the body impact angle with a simple practical paradigm ,  I showed that  current blunt impact ITBI protection standards which utilize drop tests to compare with peak acceleration or HIC apply only for  a body impact angle of $\pi/2$ (a horizonal fall to the ground). A  correction to include the effective mass of impact for  a 25\% deviation from this impact angle requires a factor of $\sim 4$ drop in the acceleration threshold for injury.
The calculations also show however, that the correction need not  much exceed this factor of 4  to accommodate almost the full range of effective impact angles.

The presentation has been minimalist  to illustrate the key ideas. More detailed development and application of these concepts  is warranted as the practical
payoffs for ITBI protection are likely to be substantial.

The HIC in its current form does not accomodate variations in surface area or effective mass.
The correction for mass  is most important 
as it   varies  substantially from  body impact orientation and its absence
 may help explain why e.g. NFL football TBI injuries are found even for lower values of the HIC than expected based on the JHTC curve (Viano et al. 2006).  The JHTC was based on moving plate impacts onto seated  monkeys  with bodies restrained.  The threshold acceleration derived from these experiments may be appropriate  when the body is supported as in a car crash, but applies to e.g. a football impact at most only for the case in which the effective mass is that of the head + helmet. The latter would be the case only when the body is strictly perpendicular to the direction of impact.
In general, injury acceleration thresholds inferred from JHTC type
 data can dangerously exceed relevant injury thresholds.


\ni {\bf Acknowledgments:} 
EGB acknowledges  the Defense Science Study Group (DSSG) 2006-2007 of the Institute for Defense Analyses for supporting  related work on TBI, and discussions  with 
Melina Hale,  Sarah Lisanby, William Moss, and 
Michael King.

\eject
\clearpage

\begin{figure}[htbp]
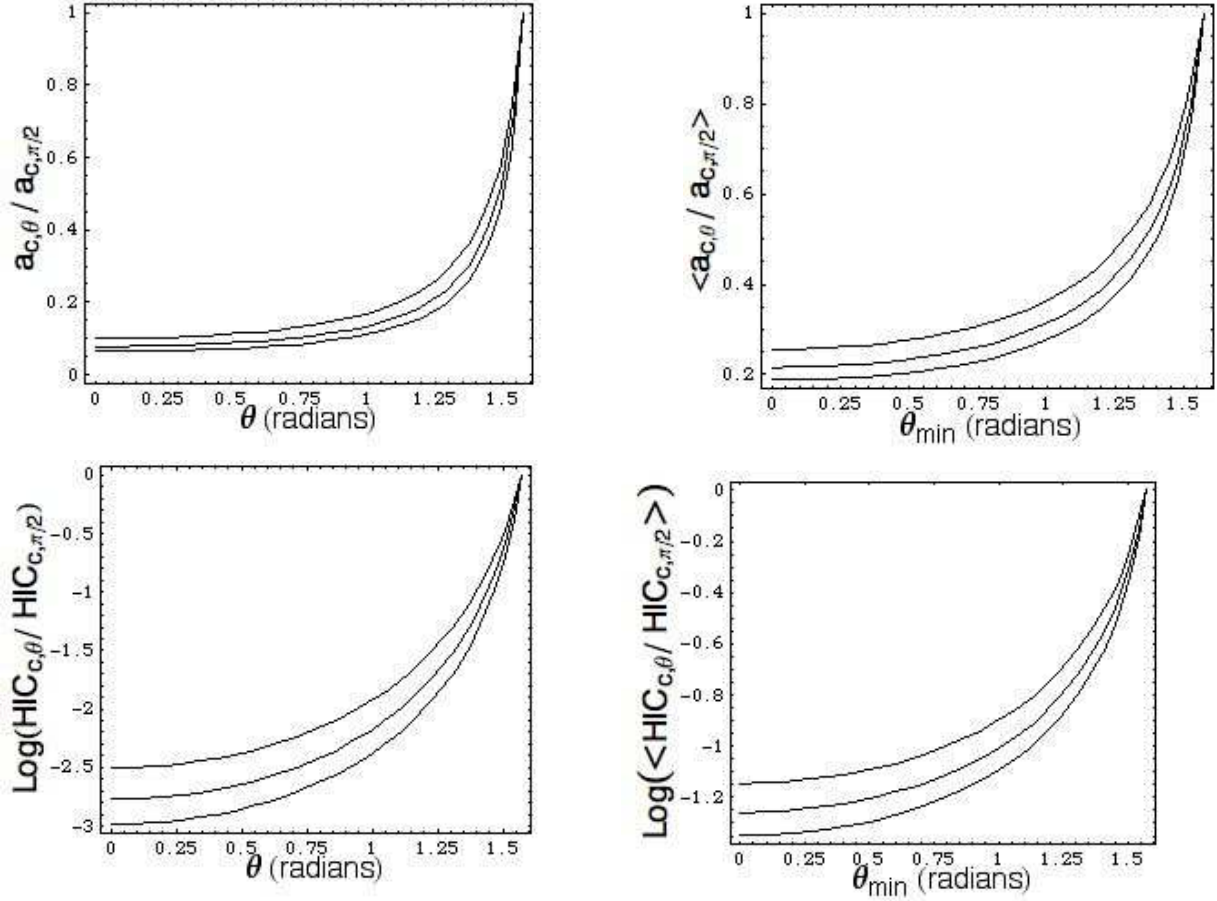

\plottwo{helmetfiga}{helmetfigb}
\plottwo{helmetfigc2}{helmetfigd2}
\caption{Left panels: Threshold acceleration and HIC for impact induced
TBI as function of  impact angle between body and impact direction 
normalized to  thresholds for an imact angle of 90 degrees.
Right panels: Threshold acceleration and HIC averaged over angle from $\theta=\theta_{min}$ to $\theta = \pi/2$ plotted vs. $\theta_{min}$.
In each panel the three curves represent, from top to bottom, a ratio of
body to head+helmet mass of  $m_b/m_h= 12.8,10, 15.6$ respectively. Larger
body to head mass means a lower acceleration threshold for injury for
all angles execpt $\pi/2$ for a fixed $m_h$.}
\end{figure}
\eject

\end{document}